\documentclass[aps,prb,superscriptaddress,twocolumn,showpacs]{revtex4}
\usepackage{dcolumn}
\usepackage{bm}
\usepackage[dvips]{graphics}
\begin{document}
\title{Off-diagonal Long-Range Order and Supersolidity
in a Quantum Solid with Vacancies}
\author{Yu Shi}
\thanks{Email address: yushi@fudan.edu.cn}
\affiliation{Department of Physics, Fudan University, Shanghai
200433, China}
\author{Yin Yang} \affiliation{School of Mathematics and Statistics,
Huazhong University of Science and Technology, Wuhan 430074, China}
\author{Shao-Ming Fei}
\affiliation{School of Mathematical Sciences, Capital Normal
University, Beijing 100048, China}
\begin{abstract}
We consider a lattice of bosonic atoms, whose number $N$ may be
smaller than the number of lattice sites  $M$. We study the
Hartree-Fock wave function built up from localized wave functions
$w(\mathbf{r})$ of single atoms, with nearest neighboring overlap.
The zero-momentum particle number is expressed in terms of
permanents of matrices. In one dimension, it is analytically
calculated to be $\alpha N(M-N+1)/M$, with $\alpha=|\int
w(\mathbf{r})d\Omega|^2/[(1+2a)l]$, where $a$ is the
nearest-neighboring overlap, $l$ is the lattice constant. $\alpha$
is of the order of $1$. The result indicates that the condensate
fraction is proportional to and of the same order of magnitude as
that of the vacancy concentration, hence there is off-diagonal
long-range order or Bose-Einstein condensation of atoms when the
number of vacancies $M-N$ is a finite fraction of the number of the
lattice sites $M$.
\end{abstract}


\maketitle

\section{Introduction}

Supersolidity refers to the superfluid-like behavior of a solid, in
particular, the non-classical rotational inertia (NCRI) or missing
moment of inertia, as a consequence of Bose-Einstein condensation
(BEC)  or off-diagonal long-range order
(ODLRO)~\cite{leggettbook,leggett,leg,shi}.  A few years ago, Kim
and Chan observed NCRI  in bulk solid $^4$He in torsional
oscillators~\cite{kim}, which  was subsequently confirmed by several
other experimental groups~\cite{reppy,kojima,kubota,shirahama,hunt}.
Heat capacity exhibits a peak near the onset of NCRI~\cite{lin}.
Superfluid-like mass flow  was seen close to the melting
temperature~\cite{day2}, and on melting curve~\cite{sasaki}, being
carried by liquid regions at the interface~\cite{sasaki2}. Recently,
it has also been observed off the melting curve by injecting atoms
from superfluids~\cite{ray}. Increase of shear modulus was observed
at low temperatures~\cite{shear}, but similar phenomenon in solid
$^3$He is not accompanied by NCRI, indicating that elastic
stiffening alone cannot produce NCRI~\cite{west}.

BEC of zero-point vacancies in the ground state of solid Helium is
the basis in some proposals of supersolidity
mechanism~\cite{andreev,chester,dai}. Path integral Monte Carlo
studies indeed found that solid $^4$He is commensurate without
BEC~\cite{ceperley}.  It was argued that zero-point vacancies or
interstitials are necessary for supersolidity~\cite{prokovev}.
Analytical calculations based on insulator-like trial wave functions
also showed that a commensurate solid cannot be a
supersolid~\cite{matsuda,imry,schwartz,shi}. But on the other hand,
zero point vacancies are found in variational studies using Jastrow
or Shadow wave function~\cite{rossi}, vacancy induced BEC was also
found by using shadow wave functions~\cite{galli}. A recent
diffusion Monte Carlo study of commensurate solid $^4$He using
another trial wave function found a condensate fraction $\sim
10^{-4}$ and a superfluid fraction $< 10^{-5}$~\cite{cazorla}.

Vacancy-based mechanism was disfavored by some researchers for the
reasons that $^4$He is believed to be commensurate while the
vacancies tend to be phase separated because of
attraction~\cite{bonin,ma,prokofev}. Nevertheless, the interaction
between vacancies may be more complicated~\cite{mahan}. On the other
hand, disorders such as dislocations, grain boundary and glassiness
indeed appear to be
important~\cite{rittner2,sasaki,hunt,kim2,prokofev,balibar}. Grain
boundaries does not seem to be the fundamental origin of NCRI, which
has also been observed in large crystals~\cite{clark}. In
considering disorder or glassiness, there are theories combining
this aspect with superfluidity~\cite{bo,toner,huse,wu,biroli}, as
well as theories without resorting to
superfluidity~\cite{nussinov,andreev2}.

With all these results, the issue whether the ground state of solid
$^4$He is commensurate or incommensurate and the mechanism of NCRI
in solid $^4$He are still open questions~\cite{gallireview}. It is
possible that intrinsic zero point vacancy is the fundamental origin
of supersolidity, while assisted by the extrinsic disorders. This
possibility is consistent with the finding in simulations that the
gap for vacancy creation can be closed under a moderate
stress~\cite{pollet}. Most recently, Anderson put forward a
Gross-Pitaevskii theory of dilute gas of vacancies to account for
the supersolidity, arguing that every pure Bose solid's ground state
is a supersolid  based on vacancies~\cite{anderson}.

As a theoretical approach shedding light on supersolid mechanism, it
is interesting to consider phenomenological trial wave functions of
a quantum solid, and examine whether they give rise to ODLRO and
supersolidity.  One of the trial wave functions is metal-like, which
is a product of copies of the same extended single atom wave
function, each being a superposition of localized wave functions at
all lattice sites. This is a BEC state, even in the case of a
perfect crystal. First studied in 1970s, this wave function suffers
the shortcoming that the probability amplitude of a configuration
with one particle on each site tends to vanish when $N \rightarrow
\infty$~\cite{imry}. Recently, it was reconsidered with
multiplication of Jastrow factors, which suppress multiple occupancy
in a same site~\cite{zhai}. However, it still has the shortcoming
that the equality between the lattice site and the number of atoms
is a coincidence~\cite{prokofev}. In the trial wave function used in
the recent diffusion Monte Carlo study which found
BEC~\cite{cazorla}, the single particle part is replaced as a
product of wave functions on all lattice sites, each being
superposition of wave functions of all possible single occupations
of this site.

Another trial wave function is insulator-like, with the
single-particle part being a symmetrized product of the localized
single-atom wave functions. There is no ODLRO in such a wave
function, even though there is wave function overlap between nearest
neighboring atoms~\cite{matsuda}. The nonexistence of ODLRO was
further proved in the cases that two particles cannot come too
close~\cite{imry}, and that the sum of overlap integrals of a single
atom wave function with its neighboring ones is less than
unity~\cite{schwartz}. Recently, the nonexistence of ODLRO or NCRI
was generally shown for the case that the overlap between the
neighboring atoms decays exponentially or faster, with the decay
constant much smaller than the system size, with or without Jastrow
factors~\cite{shi}.

On the basis of  the insulator-like wave function, Imry and Schwartz
introduced vacancies in the case that there is no overlap between
single atom local wave functions~\cite{imry}. They found the
zero-momentum particle number to be
\begin{equation}
N_0= N(M-N+1) \frac{|\int w(\mathbf{r})d\Omega|^2}{\Omega},
\label{im}
\end{equation} where $M$ is the number of total lattice sites, $N$
is the number of atoms, $w(\mathbf{r})$ is the single atom wave
function, $\Omega$ is the volume. $M \approx L^d$, $\Omega \approx M
l^d$,  where $L$ is the the number of atoms on each side of the
lattice, $d$ is the dimension, $l$ is the lattice constant.

But nearest neighboring overlap is crucial in a quantum solid.
Moreover, there is some inconsistency in discussing BEC under the
assumption that there is no overlap between neighboring atomic wave
functions. If the overlap is zero, $|\int w(\mathbf{r})d\Omega|^2$
also becomes zero. Then (\ref{im}) becomes not useful, as $N_0=0$.
To see this clearly, one can fiducially assume the single-atom wave
function to be Gaussian, as indeed used in variational calculations
of solid $^4$He~\cite{nosanow,glyde}, i.e.
\begin{equation}
w(\mathbf{r})=\frac{1}{(\sqrt{\pi} \xi)^{d/2}} \exp[-\frac{1}{2}
(\frac{r}{\xi})^2], \label{gaussian}
\end{equation} where
$d$ is the dimension of the lattice. Then
\begin{equation}
|\int w(\mathbf{r})d\Omega|^2=(2\sqrt{\pi}\xi)^{d}. \end{equation}
The nearest neighboring overlap is \begin{equation} a= \int
w(\mathbf{r})w(\mathbf{r}-\mathbf{l}) d\Omega = \exp
[-\frac{1}{4}(\frac{l}{\xi})^d].\end{equation}  Therefore the
overlap $a \rightarrow 0$ means $\xi \rightarrow 0$, or $\xi \ll l$.
But then $|\int w(\mathbf{r})d\Omega|^2 \rightarrow 0$, or $|\int
w(\mathbf{r})d\Omega|^2/l^d \ll 1$. Therefore, it is indispensable
to consider nearest neighboring overlap.

In this article, we consider the Hartree-Fock wave function of a
quantum solid with vacancies in presence of nearest neighboring
overlap of single atom wave functions. We obtain an analytical
expression for the zero-momentum particle number $N_0$, in terms of
permanents of matrices. This expression formally reduces to
Eq.~(\ref{im}) if the nearest neighboring overlap integral $a$ is
set to be $0$. We have made the analytical calculation of $N_0$ in
one dimension. Our result on $N_0$ indicates that there is ODLRO
when the number of vacancies is a finite fraction of the number of
lattice sites, in presence of nearest neighboring overlap between
single-atom wave functions.

The rest of this article is organized as follows. In Sec. II, we
consider the Hartree-Fock wave function for a quantum solid with
vacancies, constructed in terms of localized single-atom wave
functions. We obtain the analytical expression for the zero-momentum
particle number, which is expressed in terms of the permanents of
matrices. In Sec. III, we make a calculation in the case that the
overlap integral between neighboring atoms is zero, reproducing the
formula obtained by Imry and Schwartz. In Sec. III, we make the
calculation for the case that the overlap integral is nonzero. Part
of the mathematical derivation is presented in the Appendix. The
summary and discussions are made in Sec. IV.

\section{Trial wave function and the expression for zero-momentum
particle number in terms of permanents of matrices}

We consider the following Hartree-Fock wave function of a bosonic
solid with vacancies,
\begin{equation}
\Psi (\mathbf{r}_1\cdots \mathbf{r}_N)= {\cal A} \sum_I
\sum_{P_I}\prod_{i=1}^N w[\mathbf{r}_i-P_I(\mathbf{R}_{i})],
\label{psi}
\end{equation}
where $w$ is the localized single-atom wave function, which is real
and nonnegative,  $I$ represents a selection of $N$ sites
$\{\mathbf{R}_{I_1}\cdots \mathbf{R}_{I_N}\}$  from the total $M$
sites, $P_I$ represents the $N!$ permutations of these selected $N$
sites, the summation over $I$ represents $M!/N!(M-N)!$ different
choices of the $N$ sites. The normalization constant ${\cal A}$ is
obtained as
\begin{equation}
{\cal A}^{-2} = \sum_I\sum_{I'}\sum_{P_I}\sum_{P_{I'}} \prod_i
Q[P_I(\mathbf{R}_{I_i})-P_{I'}(\mathbf{R}_{I'_i})],
\end{equation}
where
\begin{equation}
Q(\mathbf{R}-\mathbf{R}') \equiv \int
w(\mathbf{r}-\mathbf{R})w(\mathbf{r}-\mathbf{R}') d\Omega. \label{q}
\end{equation}

In our consideration, \begin{equation}Q(\mathbf{R}-\mathbf{R}') =
\left\{
\begin{array}{r@{\quad: \quad {\rm if} \quad}l}
1 & \mathbf{R}-\mathbf{R}'=0, \\
a & |\mathbf{R}-\mathbf{R}'|=l,\\
0 & |\mathbf{R}-\mathbf{R}'| > l. \end{array} \right.
\end{equation}

${\cal A}^{-2}$ can be rewritten as
\begin{equation} {\cal A}^{-2}
=N!\sum_I\sum_{I'} {\cal P}[\Delta(I,I')],\end{equation} where
${\cal P}[\Delta(I,I')]$ is the permanent of an $N\times N$
submatrix $\Delta(I,I')$ of the $M\times M$ matrix ${\cal Q}$, whose
elements are
$$Q_{ij}\equiv Q(\mathbf{R}_i-\mathbf{R}_j)$$ where
$\mathbf{R}_i$ and $\mathbf{R}_j$ run over all the lattice sites.
The submatrix $\Delta(I,I')$ is formed by choosing, from ${\cal Q}$,
$N$ rows according to the set $I$ and $N$ columns according to the
set $I'$.

The permanent of an $N\times N$ matrix $\Delta$ is defined as
\begin{equation} {\cal P}(\Delta) = \sum_{i_1\cdots i_N}
\epsilon^{i_1\cdots i_N} \Delta_{1 i_1}\cdots \Delta_{N
i_N},\end{equation} where $\epsilon^{i_1\cdots i_N}=1$ when every
two indices are different from each other, otherwise
$\epsilon^{i_1\cdots i_N}=0$. In other words, a permanent is like a
determinant, except that  all the terms in the expansion are
positive, rather than with a sign alternation.

In this article, the calculation is limited to a one-dimensional
lattice. The lattice sites are numbered from left to right as $1, 2,
\cdots, M$. ${\cal Q}$ is trigonal with ${\cal Q}_{ii}=1$, ${\cal
Q}_{i,i+1}={\cal Q}_{i+1,i}=a$, while the other elements are $0$.
Therefore ${\cal Q}$ is
$${\cal Q}= A_{M};$$
here we introduce a square matrix $A_n$, $n$ being a positive
integer, written schematically as
\begin{widetext}
\begin{equation} A_{n} \equiv \left(
\begin{array}{cccccccccc}
 1&a&0&0&0&0&0&0&0&0\\
 a&1&a&0&0&0&0&0&0&0\\
 0&a&1&a&0&0&0&0&0&0\\
 0&0&a&1&a&0&0&0&0&0\\
 0&0&0&a&\ddots&\ddots&0&0&0&0\\
 0&0&0&0&a&\ddots&\ddots&0&0&0\\
 0&0&0&0&0&a&1&a&0&0\\
 0&0&0&0&0&0&a&1&a&0\\
 0&0&0&0&0&0&0&a&1&a\\
 0&0&0&0&0&0&0&0&a&1
\end{array}
\right)_{n\times n}, \label{an}
\end{equation}
where the subscript $n\times n$ indicates that it is an $n\times n$
matrix.

For the many-body trial wave function (\ref{psi}), the one-particle
reduced density matrix is
\begin{eqnarray}
\rho(\mathbf{r},\mathbf{r}') & = & N \int
\Psi(\mathbf{r},\mathbf{r}_2,
\cdots,\mathbf{r}_N)\Psi(\mathbf{r}',\mathbf{r}_2,
\cdots,\mathbf{r}_N) d\Omega_2\cdots d\Omega_N \nonumber \\
& = & N {\cal A}^2 \sum_I\sum_{I'}\sum_{P_I}\sum_{P_{I'}}
w[\mathbf{r}-P_I(\mathbf{R}_{I_1})]
w[\mathbf{r}'-P_{I'}(\mathbf{R}_{I'_1})]\prod_{i\neq 1}
Q[P_I(\mathbf{R}_{I_i})-P_{I'}(\mathbf{R}_{I'_i})] \nonumber \\
&=&{\cal A}^2 N! \sum_I\sum_{I'}\sum_{i\in I, j\in I'}
w(\mathbf{r}-\tilde{\mathbf{R}}_i)
w(\mathbf{r}'-\tilde{\mathbf{R}}_j) W_{ij},
\end{eqnarray}
\end{widetext}
where the summation over $i$ and $j$ run over the rows and columns
of submatrix $\Delta(I,I')$, $\tilde{\mathbf{R}}_i\equiv
P_I(\mathbf{R}_{I_1})$, $\tilde{\mathbf{R}}_j\equiv
P_{I'}(\mathbf{R}_{I'_1})$, for which
$Q(\tilde{\mathbf{R}}_i-\tilde{\mathbf{R}}_j)$ is just the
$(i,j)$-th element $\Delta_{ij}$ of the submatrix $\Delta(I,I')$,
$W_{ij}$ is the minor of $\Delta_{ij}$. Here the minor $W_{ij}$ of
$\Delta_{ij}$ is defined as the permanent of the submatrix of
$\Delta$ obtained by removing the $i$-th row and the $j$-th column.

The number of particles at the zero momentum state is thus
\begin{eqnarray}
N_0 & = & \frac{1}{\Omega} \int \rho(\mathbf{r},\mathbf{r}') d\Omega
d\Omega' \nonumber \\
&=& \frac{X_N({\cal Q})}{Y_N({\cal Q})} \frac{|\int w(\mathbf{r})
d\Omega|^2}{\Omega}, \label{n0}
\end{eqnarray}
where
\begin{equation}X_N({\cal Q}) \equiv \sum_I\sum_{I'}\sum_{i\in I, j\in I'}
W_{ij}, \end{equation}
\begin{equation}Y_N({\cal Q}) \equiv \sum_I\sum_{I'}{\cal P}[\Delta(I,I')].\end{equation}
$X_N({\cal Q})$ is the summation of the permanants of the minors of
all the elements of all the $N \times N$ submatrices $\Delta$'s of
${\cal Q}$. $Y_N({\cal Q})$ is the summation of the permanents of
all the $N \times N$ submatrices $\Delta$'s of ${\cal Q}$.

\section{The case without nearest neighboring overlap}

First let us reconsider the case without overlap between neighboring
single-atom wave functions, i.e. $a =0$, and show that
Eq.~(\ref{n0}) formally reduces to Eq.~(\ref{im}).

In this case,
$${\cal Q}={\cal I}_{M},$$  where ${\cal I}_{M}$ represents the
$M\times M$ unit matrix. Thus in obtaining a submatrix
$\Delta(I,I')$, once $N$ rows are chosen, there is only one choice
of $N$ columns to give rise to a nonvanishing permanent. Namely, the
ordering numbers, in the parent matrix ${\cal I}_{M}$, of the chosen
columns must be equal to those of the chosen rows, i.e.
$\Delta(I,I')$ must be a unit matrix in order to have nonvanishing
permanent. Consequently,
\begin{equation}
Y_N({\cal I}_{M}) =
\frac{M!}{N!(M-N)!},\label{denominator}\end{equation} which is just
the number of ways of choosing $N$ rows. Note that the order of the
chosen rows and the order of the chosen columns both remain the same
as in the parent matrix.

In order to calculate $X_N({\cal I}_{M})$, we need to find out all
nonzero minors for all submatrices $\Delta$'s of ${\cal I}_{M}$.
Note that ${\cal P}(\Delta)=0$ does not mean  $\Delta$ has no
nonzero minors. Given that the parent matrix is a unit matrix ${\cal
I}_{M}$, in order that $\Delta$ has one or more nonzero minors,
$\Delta$ must be either a unit matrix  ${\cal I}_{N}$ or diagonal
with only one ``$0$'' diagonal element. In the former case, there
are $M!/N!(M-N)!$ ways of making up the unit submatrix $\Delta={\cal
I}_{N}$, which has  $N$ nonzero minors, each equal to $1$. In the
latter case,  one first choose $N-1$ rows and $N-1$ columns, with
the same ordering numbers in the parent matrix ${\cal I}_{N}$,  to
make up $N-1$ diagonal elements ``1''. The number of ways of doing
this is $M!/(N-1)!(M-N+1)!$. To choose the remaining one row and one
column, their ordering numbers in the parent matrix ${\cal I}_{N}$
must be different, such that the remaining diagonal element in
$\Delta$ is ``0''. The number of ways of doing this is
$(M-N+1)(M-N)$. Each $\Delta$ so obtained only has one nonzero
minor, which is equal to $1$. Therefore, \begin{widetext}
\begin{equation}
\begin{array}{rcl}
X_N({\cal I}_{M}) & = &\frac{M!}{N!(M-N)!} N
+\frac{M!}{(N-1)!(M-N+1)!} (M-N)(M-N+1) \nonumber
\\
&=& \frac{M!}{N!(M-N)!} N(M-N+1).\label{numerator}\end{array}
\end{equation}
\end{widetext}

Substituting (\ref{denominator}) and (\ref{numerator}) into
Eq.~(\ref{n0}) indeed recovers Eq.~(\ref{im}).

\section{The case with nearest neighboring overlap}

With nearest neighboring overlap,  the zero-momentum particle number
is
\begin{equation}
N_0 = \frac{X_N(A_{M})}{Y_N(A_{M})} \frac{|\int w(\mathbf{r})
d\Omega|^2}{\Omega}, \label{nl}
\end{equation}
where the matrix $A_M$ is as defined in Eq.~(\ref{an}).

For an arbitrary matrix $S_{m\times n}$, we introduce
$Y_k(S_{m\times n})$ and $X_k(S_{m\times n})$, with $k\leq \min
(m,n)$. $Y_k(S_{m\times n})$ is the sum of the permanents of all the
$k\times k$ submatrices of $S_{m\times n}$. $X_k(S_{m\times n})$ is
the sum of the permanents of all the minors of all the $k\times k$
submatrices of $S_{m\times n}$.

First, we note the existence of the relation
\begin{equation} X_k(A_n) = (n-k+1)^2 Y_{k-1}(A_n), \label{rel}
\end{equation} for the following reason. Every minor of a  $k\times
k$ submatrix of $A_n$ is in fact a $(k-1) \times (k-1)$ submatrix of
$A_n$, while a $(k-1) \times (k-1)$ submatrix is a minor of many
different $k\times k$ submatrices. For a given $(k-1) \times (k-1)$
submatrix of $A_n$, one can add an additional row and an additional
column of the $A_n$, making up a $k\times k$ submatrix of $A_n$, of
which the concerned $(k-1) \times (k-1)$ submatrix of $A_n$ is a
minor. There are $(n-k+1)^2$ ways to do this. Hence a $(k-1) \times
(k-1)$ submatrix is a minor of $(n-k+1)^2$ different $k\times k$
submatrices of $A_n$, thus we obtain the relation (\ref{rel}).

Therefore
\begin{equation} \frac{X_k(A_n)}{Y_k(A_n)}=\frac{(n-k+1)^2
Y_{k-1}(A_n)}{Y_k(A_n)}. \label{ratio}\end{equation}

In Appendix A, we obtain that for $n \geq 2$,
\begin{equation}
\begin{array}{rl}
Y_2(A_n)& =(1+2a)^2\displaystyle\frac{n(n-1)}{2}-(5a^2+4a)n+7a^2+4a \nonumber \\
& =(1+2a)^2\displaystyle\frac{n(n-1)}{2!}[1+O(\frac{1}{n})],
\end{array}\end{equation} where $O(1/n)$ represents a term of the order of
$1/n$.

In Appendix B, we obtain that for any  $ 3\leq k < n,$
\begin{widetext}
\begin{eqnarray}
Y_k(A_n)& = & \sum_{l=k-1}^{n-1}Y_{k-1}(A_{l})
+2\sum_{s=1}^{k-2}a^s\sum_{l=k-s}^{n-s-1}Y_{k-s}(A_{l})+a^2\sum_{l=k-2}^{n-2}
Y_{k-2}(A_{l}) \nonumber
\\
&& +(1+2a)2a^{k-1}(n-k)(n-k+1). \label{rel2}
\end{eqnarray}
\end{widetext}

From this relation,  we know that $Y_k(A_n) > Y_{k-1} (A_{n-1})$,
$Y_{k-1} (A_{n-1})$ being merely one term in the first summation in
RHS of (\ref{rel2}). Consequently, in the summation over $s$,
$2\sum_{l=k-s}^{n-s-1}Y_{k-s}(A_{l})$, which also depends on $a$, by
which $a^s$ is multiplied, decreases with the increase of $s$. Since
$a<1$, RHS of (\ref{rel2}) converges with respect to $a$. Also note
that the last term is of the power of $a^{k-1}$.

Therefore, $Y_k(A_n)$ can be written as
\begin{widetext}
\begin{eqnarray} Y_k(A_n) & = &[\sum_{l=k-1}^{n-1}Y_{k-1}(A_{l})+
2a\sum_{l=k-1}^{n-2}Y_{k-1}(A_{l})][1+O_{\leq}(a)] \nonumber \\
& = & [(1+2a)\sum_{l=k-1}^{n-1}Y_{k-1}(A_{l})-2a
Y_{k-1}(A_{n-1})][1+O_{\leq}(a)], \label{ykan}
\end{eqnarray}
\end{widetext} where $O_{\leq}(a)$ denotes a term at most of the order
of $a$.

In the following, we show by induction that \begin{equation}
Y_k(A_n) = (1+2a)^k \frac{n!}{k!(n-k)!}[1+O_{\leq}(a)]. \label{yk}
\end{equation}

Suppose that the similar identity is valid for $Y_{k-1}(A_l)$, with
$k-1 \leq l <n$, i.e.,
\begin{equation}  Y_{k-1}(A_l) = (1+2a)^{k-1}
\frac{l!}{(k-1)!(l-k+1)!}[1+O_{\leq}(a)].
\label{suppose}\end{equation}

Then \begin{widetext}\begin{eqnarray}
(1+2a)\sum_{l=k-1}^{n-1}Y_{k-1}(A_{l}) & =& \frac{(1+2a)^k}{(k-1)!}
\sum_{l=k-1}^{n-1} \frac{l!}{(l-k+1)!}[1+O_{\leq}(a)]. \label{fg}
\end{eqnarray}\end{widetext}

Using the identity~\cite{identity} \begin{equation} \sum_{j=1}^p
j(j+1)\cdots (j+q) = \frac{1}{q+2}\frac{(p+q+1)!}{(p-1)!},
\label{identity1}
\end{equation} we obtain
\begin{equation}\sum_{l=k-1}^{n-1}
\frac{l!}{(l-k+1)!}=\frac{1}{k}\frac{n!}{(n-k)!},\end{equation}
Hence (\ref{fg}) becomes
\begin{equation} (1+2a)\sum_{l=k-1}^{n-1}Y_{k-1}(A_{l})
=(1+2a)^k\frac{n!}{k!(n-k)!}[1+O_{\leq}(a)]. \label{s1}
\end{equation}

On the other hand, according to the assumption (\ref{suppose}) and
the above result (\ref{s1}), we have
\begin{widetext}
\begin{eqnarray}
2a Y_{k-1}(A_{n-1}) & = & 2a
(1+2a)^{k-1}\frac{(n-1)!}{(k-1)!(n-k)!} [1+O_{\leq}(a)] \nonumber \\
& = & \frac{2a}{1+2a}\frac{k}{n}[(1+2a)
\sum_{l=k-1}^{n-1}Y_{k-1}(A_{l})][1+O_{\leq}(a)]
\nonumber \\
& = & O(a) \frac{k}{n} \sum_{l=k-1}^{n-1}Y_{k-1}(A_{l}), \label{s2}
\end{eqnarray}\end{widetext}
where $k/n \leq 1 $.

Substituting (\ref{s1}) and (\ref{s2}) into Eq.~(\ref{ykan}) yields
Eq.~(\ref{yk}), hence the proof completes.

Then we substitute the proved identity (\ref{yk}) into
Eq.~(\ref{ratio}),  obtaining

\begin{equation}
\frac{X_k(A_n)}{Y_k(A_n)}\approx \frac{k(n-k+1)}{1+2a},
\label{ratio2}\end{equation} if $a \ll 1$.

Therefore, according to (\ref{nl}),  the zero-momentum particle
number is
\begin{equation}
N_0 \approx \frac{N(M-N+1)}{1+2a} \frac{|\int w(\mathbf{r})
d\Omega|^2}{\Omega}. \label{n2}
\end{equation}
This identity formally reduces to that obtained by Imry and Schwartz
long ago when we set $a=0$.  But $|\int w(\mathbf{r})d\Omega|^2 \neq
0$ only if $a \neq 0$.

On the other hand, in case $n \gg k$, no matter whether $a \ll 1$,
we can also obtain the identity (\ref{ratio2}) and thus the result
(\ref{n2}). This condition does not correspond to the physical
situation concerning solid $^4$He, as that would mean most of the
lattice sites are empty. For completeness, we give the mathematical
proof for this case in Appendix C.

\section{Summary and Discussions}

To summarize, we have studied the Hartree-Fock wave function of a
lattice of atoms with $N \leq M$, where  $M$ and $N$ are the numbers
of lattice sites and atoms, respectively. The Hartree-Fock wave
function is constructed in terms of localized wave functions of
single atoms, with nearest neighboring overlap.

In one dimension, under this wave function, we have obtained the
zero-momentum particle number as given in (\ref{n2}), which can be
rewritten as
\begin{equation} N_0 = \alpha \frac{N(M-N+1)}{M},
\label{eq} \end{equation} where
\begin{equation}\alpha=\frac{|\int
w(\mathbf{r})d\Omega|^2}{(1+2a)l}\end{equation} is a finite fraction
of the order of $1$.

To be specific, let us again use the Gaussian wave function for
$w(\mathbf{r})$, as given in (\ref{gaussian}). Then in one
dimension,
\begin{equation}\alpha = \frac{2\sqrt{\pi}\frac{\xi}{l}}{[1+2\exp (-\frac{1}{4}
(\frac{l}{\xi})^2]},\end{equation} which is of the order of $1$ when
$\xi/l$ is a finite fraction around  $0.36$, which is obtained from
the Lindemann ratio $\delta\approx 0.29$ for solid
$^4$He~\cite{glyde}, using $\delta\equiv \sqrt{\langle r^2 \rangle
}/l = \sqrt{3/2} \xi/l$ under the Gaussion wave function
(\ref{gaussian}). For $d=3$, $1/(1+2a)$ should be replaced by
another function $f(\alpha)$ of $\alpha$,  which should still be of
the order of $1$. Anyway, $\alpha=(2\sqrt{\pi}\xi/l)^3f(\alpha)$
must be of the order of $1$.

Therefore there is BEC of atoms, i.e. $N_0$ is a finite fraction of
$N$, when the number of vacancies $M-N$ is a finite fraction of the
number of lattice sites $M$. This condition also implies that the
number of atoms $N$ is a finite fraction of $M$.

Interestingly, the condensate fraction $N_0/N$ is proportional to
and of the order of  vacancy concentration $(M-N)/M$,
\begin{equation}\frac{N_0}{N} = \alpha \frac{M-N}{M}.\end{equation} Currently, the
experimental upper bound of vacancy concentration is about
$0.4\%$~\cite{rossi}.   Hence a Hartree-Fock wave function for a
solid with zero point vacancy implies that the condensate fraction
is about $0.004\alpha$, which is very reasonable.

Moreover, for such low vacancy concentration, one has
\begin{equation} \frac{N_0}{M-N} \approx \alpha ,\end{equation} i.e. $\alpha$ equals the
number of condensed atoms per vacancy. This is well consistent with
the result of variational simulation based on Shadow wave function,
which gives $0.23$ condensed atoms per vacancy at 54
bar~\cite{galli}.

The Hartree-Fock wave function could be the ground state of a mean
field theory.  Although it is not multiplied by the Jastrow factor,
the double occupancy is excluded by construction. Our calculation is
done for one dimension.  In three dimensions, there should not be
qualitative difference in order of magnitude from the result for one
dimension. Hence our result is qualitatively informative for solid
$^4$He, suggesting that its supersolidity based on BEC of atoms
induced by zero point vacancy is possible.

\acknowledgments

We thank A. J. Leggett, X. Q. Li-Jost, L. Reatto and Y. S. Wu for
useful discussions. This work is supported by National Science
Foundation of China (Grant No. 10674030), Shuguang Project (Grant
No. 07S402) and The Ministry of Science and Technology of China
(Grant No. 2009CB929204).

\appendix

\begin{widetext}

\section{$Y_2(A_n)$}

Here we calculate $Y_2(A_n)$ ($n\geq 2$), i.e. the sum of the
permanents of all the $2\times 2$ submatrices of $A_n$.

In addition to the definition of $A_n$ as given in (\ref{an}), we
shall also use matrices
\begin{equation} B_{n-1} \equiv \left(
\begin{array}{ccccccccc}
 a&a&0&0&0&0&0&0&0\\
 0&1&a&0&0&0&0&0&0\\
 0&a&1&a&0&0&0&0&0\\
 0&0&a&1&a&0&0&0&0\\
 0&0&0&\ddots&\ddots&0&0&0&0\\
 0&0&0&0&\ddots&\ddots&a&0&0\\
 0&0&0&0&0&a&1&a&0\\
 0&0&0&0&0&0&a&1&a\\
 0&0&0&0&0&0&0&a&1
\end{array}
\right)_{(n-1)\times(n-1)},\end{equation}
\begin{equation}C_{(n-2)\times(n-1)} \equiv \left(
\begin{array}{ccccccccc}
 a&1&a&0&0&0&0&0&0\\
 0&a&1&a&0&0&0&0&0\\
 0&0&a&1&a&0&0&0&0\\
 0&0&0&a&1&a&0&0&0\\
 0&0&0&\ddots&\ddots&0&0&0&0\\
 0&0&0&0&\ddots&\ddots&1&a&0\\
 0&0&0&0&0&0&a&1&a\\
 0&0&0&0&0&0&0&a&1
\end{array}
\right)_{(n-2)\times(n-1)}.\end{equation}

The permanent of an $n \times n$ matrix $S$ is equal to
\begin{equation}{\cal P}(S)= \sum_{j}S_{ij}W_{ij},\end{equation} where $W_{ij}$
is the minor of the $S_{ij}$. Using this property, we can expand
$Y_2(A_n)$ as
\begin{eqnarray}
Y_2(A_n) & \equiv& Y_2 \left(
\begin{array}{cccccccccc}
 1&a&0&0&0&0&0&0&0&0\\
 a&1&a&0&0&0&0&0&0&0\\
 0&a&1&a&0&0&0&0&0&0\\
 0&0&a&1&a&0&0&0&0&0\\
 0&0&0&a&\ddots&\ddots&0&0&0&0\\
 0&0&0&0&a&\ddots&\ddots&0&0&0\\
 0&0&0&0&0&a&1&a&0&0\\
 0&0&0&0&0&0&a&1&a&0\\
 0&0&0&0&0&0&0&a&1&a\\
 0&0&0&0&0&0&0&0&a&1
\end{array}
\right)_{n\times n} \nonumber \\
&= &  Y_1(A_{n-1})+a  Y_1(B_{(n-1)\times(n-1)})
+a Y_1(C_{(n-2)\times(n-1)})+ Y_2(A_{n-1}) \label{y2an} \\
&= & Y_2(A_{n-1})+Y_1(A_{n-1})+2a Y_1(A_{n-2})+ 3a^2, \nonumber
\end{eqnarray}
where we have used $Y_1(B_{(n-1)\times(n-1)})=2a+Y_1(A_{n-2})$,
$Y_1(C_{(n-2)\times(n-1)})=a+Y_1(A_{n-2})$, which are
straightforward.

In this way, we obtain the following set of identities
\begin{equation}\begin{array}{rcl}
Y_2(A_n)-Y_2(A_{n-1})&=& Y_1(A_{n-1})+2a
Y_1(A_{(n-2)})+ 3a^2, \\
Y_2(A_{n-1})-Y_2(A_{n-2})&=& Y_1(A_{n-2})+2a
Y_1(A_{(n-3)})+ 3a^2, \\
&\vdots& \\
Y_2(A_{3})-Y_2(A_{2})&=&Y_1(A_{2})+2a Y_1(A_{1})+ 3a^2.\end{array}
\end{equation}

Adding these identities together gives rise to
\begin{equation}Y_2(A_n)-Y_2(A_{2})=Y_1(A_{n-1})+(1+2a)\sum_{j=2}^{n-2} Y_1(A_j)+2a
Y_1(A_{1})+3a^2(n-2).\end{equation} Clearly $Y_2(A_{2})=1+a^2$,
$Y_1(A_j) = j+2(j-1)a$. Hence it can be obtained that
\begin{equation}Y_2(A_n)= (1+2a)^2\frac{n(n-1)}{2} - (5a^2+4a) n +
7a^2+4a, \end{equation} which is also satisfied when $n=2$, as
$Y_2(A_2) =1 +a^2$.

\section{$ Y_k(A_n) $  }

We now calculate $Y_k(A_n)$ for $3 \leq k < n$, in a way similar to
the calculation of $Y_2(A_n)$ above.

Similar to (\ref{y2an}), we obtain
\begin{equation}
\begin{array}{rl}
 Y_k(A_n)&
 \equiv Y_k \left(
\begin{array}{cccccccccc}
 1&a&0&0&0&0&0&0&0&0\\
 a&1&a&0&0&0&0&0&0&0\\
 0&a&1&a&0&0&0&0&0&0\\
 0&0&a&1&a&0&0&0&0&0\\
 0&0&0&a&\ddots&\ddots&0&0&0&0\\
 0&0&0&0&a&\ddots&\ddots&0&0&0\\
 0&0&0&0&0&a&1&a&0&0\\
 0&0&0&0&0&0&a&1&a&0\\
 0&0&0&0&0&0&0&a&1&a\\
 0&0&0&0&0&0&0&0&a&1
\end{array}
\right)_{n\times n} \\
= &  Y_{k-1}(A_{n-1})+a Y_{k-1}(B_{(n-1)\times(n-1)})+a
Y_{k-1}(C_{(n-2)\times(n-1)})+ Y_{k}(A_{n-1}).
\end{array}
\end{equation}
Expansion of $Y_{k-1}(B_{(n-1)\times(n-1)})$ gives
\begin{equation}
Y_{k-1}(B_{(n-1)\times(n-1)}) =a
Y_{k-2}(A_{k-2})+Y_{k-1}(C^T_{(n-1)\times(n-2)}).\end{equation} For
any matrix $S$, $Y_{k-1}(S^T)= Y_{k-1}(S)$. Hence we have
\begin{equation}
Y_k(A_n) =  Y_k(A_{n-1})+Y_{k-1}(A_{n-1})+a^2Y_{k-2}(A_{n-2}) +2a
Y_{k-1}(C_{(n-2)\times(n-1)}). \label{yk1}\end{equation}

Iterative expansion of $Y_{k-1}(C_{(n-2)\times(n-1)})$ yields
\begin{equation}
\begin{array}{rcl}
Y_{k-1}(C_{(n-2)\times(n-1)})
&=&Y_{k-1}(A_{n-2})+a Y_{k-2}(C_{(n-3)\times(n-2)})  \\
&=&\cdots \\
&=&\displaystyle\sum_{s=1}^{k-2} a^{s-1}Y_{k-s}(A_{n-s-1})
+a^{k-2}Y_{1}(C_{(n-k)\times(n-k+1)}),
\end{array}\end{equation}
where $Y_{1}(C_{(n-k)\times(n-k+1)})=(n-k)+a+2(n-k-1)a$.

Therefore
\begin{equation}
\begin{array}{rcl}
Y_k(A_n)
- Y_k(A_{n-1})&=&Y_{k-1}(A_{n-1})+a^2Y_{k-2}(A_{n-2}) +2a Y_{k-1}(C_{(n-2)\times(n-1)})\\
&=&Y_{k-1}(A_{n-1})+a^2Y_{k-2}(A_{n-2}) \\
&&+2\displaystyle\sum_{s=1}^{k-2}a^sY_{k-s}(A_{n-s-1})
+2a^{k-1}[(n-k)+a+2(n-k-1)a].
\end{array}
\end{equation}

Hence
\begin{equation}
\begin{array}{rl}
Y_k(A_n) - Y_k(A_{n-1})
=&Y_{k-1}(A_{n-1})+a^2Y_{k-2}(A_{n-2})\\&+2\displaystyle\sum_{s=1}^{k-2}a^sY_{k-s}(A_{n-s-1})
+2a^{k-1}[a+(n-k)+2(n-k-1)a]\\
Y_k(A_{n-1}) - Y_k(A_{n-2})
=&Y_{k-1}(A_{n-2})+a^2Y_{k-2}(A_{n-3})\\&+2\displaystyle\sum_{s=1}^{k-2}a^sY_{k-s}(A_{n-s-2})
+2a^{k-1}[a+(n-1-k)+2(n-k-2)a]\\
\vdots & \\
Y_k(A_{k+2}) - Y_k(A_{k+1})
=&Y_{k-1}(A_{k+1})+a^2Y_{k-2}(A_{k})\\&+2\displaystyle\sum_{s=1}^{k-2}a^sY_{k-s}(A_{k+1-s})
+2a^{k-1}[a+2+2a]\}\\
Y_k(A_{k+1}) - Y_k(A_{k})
=&Y_{k-1}(A_{k})+a^2Y_{k-2}(A_{k-1})\\&+2\displaystyle\sum_{i=1}^{k-2}a^sY_{k-s}(A_{k-s})
+2a^{k-1}[a+1]\}\\
\end{array}
\end{equation}

Adding these identities together leads to
\begin{equation}
\begin{array}{rl}
Y_k(A_n)-Y_k(A_{k}) &=  \displaystyle\sum_{l=k}^{n-1}Y_{k-1}(A_{l})
+a^2\displaystyle\sum_{l=k-1}^{n-2}Y_{k-2}(A_{l})
+2\displaystyle\sum_{s=1}^{k-2}a^s\displaystyle
\sum_{l=k-s}^{n-s-1}Y_{k-s}(A_{l})\\
&+2a^{k-1}\displaystyle\sum_{s=0}^{n-k-1}[a+(n-k-s)+2(n-k-1-s)a]\\
\end{array}
\end{equation}
Since  $Y_m(A_n)$ exists only when $m\leq n$, we have
\begin{equation}
Y_k(A_k)=Y_{k-1}(A_{k-1})+a^2Y_{k-2}(A_{k-2}).
\end{equation}
Therefore
\begin{eqnarray}
Y_k(A_n)& = & \sum_{l=k-1}^{n-1}Y_{k-1}(A_{l})
+2\sum_{s=1}^{k-2}a^s\sum_{l=k-s}^{n-s-1}Y_{k-s}(A_{l})
+a^2\sum_{l=k-2}^{n-2} Y_{k-2}(A_{l}) \nonumber
\\
&&+(1+2a)2a^{k-1}(n-k)(n-k+1). \label{relb}
\end{eqnarray}

\section{Calculation of $N_0$ in the case of $n \gg k$}

Here we show that the identity (\ref{ratio2}) and thus the result
(\ref{n2}) are also valid if $n \gg k$, no matter whether $a \ll 1$
or not. Mathematically, $n \gg k$ means $n \rightarrow \infty$ while
$k$ remains finite.

In the following, we show by induction that \begin{equation}
Y_k(A_n) = (1+2a)^k \frac{n!}{k!(n-k)!}[1+O(a)O(\frac{1}{n})].
\label{yk2}
\end{equation}

Suppose that the similar identity is valid for $Y_{k-s}(A_l)$, with
$s \geq 1$ and  $k-s \leq l <n$, i.e.,
\begin{equation}  Y_{k-s}(A_l) = (1+2a)^{k-s}
\frac{l!}{(k-s)!(l-k+s)!}[1+O(a)O(\frac{1}{l})].
\label{suppose2}\end{equation}

In the second term in the exact identity (\ref{relb}) for
$Y_k(A_n)$, $2a^s$ is multiplied by
$\sum_{l=k-s}^{n-s-1}Y_{k-s}(A_{l})$, which can be evaluated by
using the assumption (\ref{suppose2}) to be
\begin{eqnarray}
\sum_{l=k-s}^{n-s-1}Y_{k-s}(A_{l}) & = &
\frac{(1+2a)^{k-s}}{(k-s)!}\displaystyle\sum_{l=k-s}^{n-s-1}\frac{l!}{(l-k+s)!}[1+O(a)O(\frac{1}{l})]
\nonumber\\
&=&\frac{(1+2a)^{k-s}}{(k-s+1)!}\frac{(n-s)!}{(n-k-1)!}
\nonumber \\
&=&[\sum_{l=k-1}^{n-2}Y_{k-1}(A_{l})]\frac{(k-2)\cdots
(k-s+1)}{(1+2a)^{s-1}} \frac{1}{(n-1)\cdots (n-s+1)}, \label{asc}
\end{eqnarray}
where we have used the identity (\ref{identity1}).

In the third term in (\ref{relb}), $2a^2$ is multiplied by
$\sum_{l=k-2}^{n-2}Y_{k-2}(A_{l})$, which can be similarly evaluated
to be
\begin{eqnarray}
\sum_{l=k-2}^{n-2}Y_{k-2}(A_{l}) & = &
\frac{(1+2a)^{k-2}}{(k-2)!}\displaystyle\sum_{l=k-2}^{n-2}\frac{l!}{(l-k+2)!}[1+O(a)O(\frac{1}{l})]
\nonumber\\&=&\frac{(1+2a)^{k-2}}{(k-1)!}\frac{(n-1)!}{(n-k)!}
\nonumber
\\
&=&[\displaystyle\sum_{l=k-1}^{n-2}Y_{k-1}(A_{l})]\frac{k}{(1+2a)}\frac{1}{(n-k)},
\label{a2c}
\end{eqnarray}
where we have also used the identity (\ref{identity1}).

Besides, the last term in (\ref{relb}) is $O(a^{k-1})O(n^2) \ll
Y_{k-1}(A_{n-1}) = O(n^k)$ if $k \ll n$.

Therefore
\begin{eqnarray} Y_k(A_n) & =
&[\sum_{l=k-1}^{n-1}Y_{k-1}(A_{l})+
2a\sum_{l=k-1}^{n-2}Y_{k-1}(A_{l})][1+O(a)O(\frac{1}{n})],
\label{ykan2}
\end{eqnarray}
where $O(1/n) \ll 1$, $O(a)$ is of the order of $a$, which we do not
need to specify.

Using the assumption (\ref{suppose2}) for $s=1$, we obtain
\begin{eqnarray} (1+2a)\sum_{l=k-1}^{n-1}Y_{k-1}(A_{l}) & =&
(1+2a)^k \sum_{l=k-1}^{n-1}
\frac{l!}{(k-1)!(l-k+1)!}[1+O(\frac{1}{l})] \nonumber
\\ & = & \frac{(1+2a)^k}{(k-1)!}[ F + O(G)], \label{fg2}
\end{eqnarray}
where \begin{equation}F=\sum_{l=k-1}^{n-1}
\frac{l!}{(l-k+1)!},\end{equation}
\begin{equation}G=\sum_{l=k-1}^{n-1} \frac{(l-1)!}{(l-k+1)!}.\end{equation}

Using the identity (\ref{identity1}),  we obtain
\begin{equation}F=\frac{1}{k}\frac{n!}{(n-k)!},\end{equation}
\begin{equation}G=\frac{1}{k-1}\frac{(n-1)!}{(n-k)!}.\end{equation}
Hence (\ref{fg2}) becomes
\begin{equation} (1+2a)\sum_{l=k-1}^{n-1}Y_{k-1}(A_{l})
=\frac{(1+2a)^k}{k!}\frac{n!}{(n-k)!}[1+O(a)O(\frac{1}{n})].
\label{s12}
\end{equation}

On the other hand, according to (\ref{suppose2}) and (\ref{s12}),
\begin{eqnarray}
2a Y_{k-1}(A_{n-1}) & = & \frac{2a
(1+2a)^{k-1}}{(k-1)!}\frac{(n-1)!}{(n-k)!}[1+O(a)O(\frac{1}{n})] \nonumber \\
& = &
\frac{2a}{1+2a}\frac{k}{n}[(1+2a)\sum_{l=k-1}^{n-1}Y_{k-1}(A_{l})]
\nonumber \\
& = & [(1+2a)\sum_{l=k-1}^{n-1}Y_{k-1}(A_{l})]O(a)O(\frac{1}{n}),
\label{s22}
\end{eqnarray}
for the reason that $k \ll n$.

Therefore,
\begin{eqnarray}
\sum_{l=k-1}^{n-1}Y_{k-1}(A_{l})+ 2a\sum_{l=k-1}^{n-2}Y_{k-1}(A_{l})
\nonumber \\
=(1+2a)\sum_{l=k-1}^{n-1}Y_{k-1}(A_{l})-2a Y_{k-1}(A_{n-1})
\nonumber
\\
=(1+2a)\sum_{l=k-1}^{n-1}Y_{k-1}(A_{l})[1+O(a)O(\frac{1}{n})].
\label{xc}
\end{eqnarray}

By using (\ref{s12}) and (\ref{xc}), (\ref{ykan2}) leads to the
identity (\ref{yk2}), which is thus proved.  Substituting this
proved identity into Eq.~(\ref{ratio}), we obtain Eq.~(\ref{ratio2})
and thus also (\ref{n2}).
\end{widetext}

\end{document}